 \documentclass[useAMS,usenatbib,usegraphicx]{mn2e}
 \usepackage{amssymb}
 \usepackage{aas_macros}
 \usepackage{amsmath}
 
 \def\iraf{\textsc{iraf}}
 \def\ccdproc{\textsc{ccdproc}}
 \def\quadproc{\textsc{quadproc}}
 \def\kpnoslit{\textsc{kpnoslit}}
 \def\rvsao{\textsc{rvsao}}
 \def\molly{\textsc{molly}}
 \def\doppler{\textsc{doppler}}
 \def\elc{\textsc{elc}}
 \def\ut{\textsc{ut}}
 \def\tdb{\textsc{tdb}}
 \def\ttime{\textsc{tt}}
 \def\mjd{\textsc{mjd}}
 \def\hjd{\textsc{hjd}}
 \def\xtime{\textsc{xTime}}
 
 \def\daophot{\textsc{daophot}}
 
 \newcommand\ion[2]{#1$\;${\scshape{#2}}}

 \title[Optical observations of SAX~J1808.4$-$3658]{Optical spectroscopy and photometry of SAX~J1808.4$-$3658 in outburst}
 \author[P. Elebert et al.]{P. Elebert,$^{1}$\thanks{E-mail: p.elebert@ucc.ie}
         M.~T.~Reynolds,$^{2}$
         P.~J.~Callanan,$^{1}$
         D.~J.~Hurley,$^{1}$
         G.~Ramsay,$^{3}$
         \newauthor F.~Lewis,$^{4,5,6}$
         D.~M.~Russell,$^{7}$
         B.~Nord,$^{8}$
         S.~R.~Kane,$^{9}$
         D.~L.~DePoy$^{10}$ and
         \newauthor P.~Hakala$^{11}$\\
 $^{1}$Department of Physics, University College Cork, Cork, Ireland\\
 $^{2}$Department of Astronomy, University of Michigan, 500 Church Street, Ann Arbor, MI 48109-1042, USA\\
 $^{3}$Armagh Observatory, College Hill, Armagh, BT61 9DG, Northern Ireland\\
 $^{4}$Faulkes Telescope Project, School of Physics and Astronomy, Cardiff University, 5, The Parade, Cardiff, CF24 3AA, Wales\\
 $^{5}$Department of Physics and Astronomy, The Open University, Walton Hall, Milton Keynes, MK7 6AA, UK\\
 $^{6}$Las Cumbres Observatory Global Telescope, 6740 Cortona Drive, Goleta, CA 93117, USA\\
 $^{7}$Astronomical Institute ``Anton Pannekoek'', Kruislaan 403, 1098 SJ Amsterdam, the Netherlands\\
 $^{8}$Physics Department, University of Michigan, 450 Church Street, Ann Arbor, MI 48109-1040, USA\\
 $^{9}$NASA Exoplanet Science Institute, Caltech, MS 100-22, 770 South Wilson Avenue, Pasadena, CA 91125, USA\\
 $^{10}$Department of Physics, Texas A\&M University, 4242 TAMU, College Station, TX 77843-4242, USA\\
 $^{11}$Tuorla Observatory, University of Turku, V\"{a}is\"{a}l\"{a}ntie 20, FIN-21500 Piikki\"{o}, Finland}
 
\begin{document}
 
 
 \pagerange{\pageref{firstpage}--\pageref{lastpage}} \pubyear{2009}
 
 \maketitle
 
 \label{firstpage}


 \begin{abstract}
     \noindent
     We present phase resolved optical spectroscopy and photometry of V4580~Sagittarii, the optical counterpart
     to the accretion powered millisecond pulsar SAX~J1808.4$-$3658, obtained during the 2008 September/October
     outburst.
     Doppler tomography of the \ion{N}{iii} $\lambda$4640.64 Bowen blend emission line
     reveals a focused spot of emission at a location consistent with the secondary star.
     The velocity of this emission occurs
     at $324 \pm 15$~km~s$^{-1}$; applying a ``$K$-correction'', we find the velocity of the
     secondary star
     projected onto the line of sight to be $370 \pm 40$~km~s$^{-1}$. Based on existing pulse
     timing measurements,
     this constrains the mass ratio of the system to be $0.044^{+0.005}_{-0.004}$, and the mass function for
     the pulsar to be $0.44^{+0.16}_{-0.13}$~M$_{\sun}$. Combining this mass function with various
     inclination estimates from other authors, we find no evidence to suggest that the neutron
     star in SAX~J1808.4$-$3658 is more massive than the canonical value of 1.4~M$_{\sun}$.
     Our optical light curves exhibit a possible superhump modulation, expected for a system with such a low
     mass ratio.
     The equivalent width of the \ion{Ca}{ii} H and K interstellar absorption lines suggest that the distance to the source 
     is $\sim$2.5~kpc. This is consistent with previous distance estimates based on type-I X-ray bursts which assume
     cosmic abundances of hydrogen, but lower than more recent estimates which assume helium-rich bursts.
 \end{abstract}
 

 \begin{keywords}
     accretion, accretion discs --
     binaries: close --
     pulsars: individual: SAX~J1808.4$-$3658 --
     stars: individual: V4580~Sagittarii --
     stars: neutron --
     X-rays: binaries
 \end{keywords}

 
 \section{Introduction}
 \label{intro}
 
 Low mass X-ray binaries (LMXBs) are systems containing a low mass secondary ($M_2
 \lesssim$ 1 M$_{\sun}$) and a compact primary, either a neutron
 star (NS) or a black hole (BH). These may be divided into two subclasses based on
 whether they are persistent or transient sources.
 The transient systems or X-ray Novae (XRNe) are typically characterised by short periods
 of heightened luminosity separated by long periods
 of quiescence. During an XRN outburst the X-ray luminosity
 increases by 10$^4$ -- 10$^6$, reaching a sizable fraction of the Eddington
 luminosity. The X-ray outburst is accompanied by an outburst at
 UV/optical/IR wavelengths dominated by reprocessing of X-rays in the
 accretion disc \citep[for example, see the review by][]{charles2006}.
 
 Until recently, the vast majority of known XRNe were black hole
 systems. As a result, systematic comparison between the accretion discs in
 BH and NS transients was difficult. At the same time,
 there was little direct evidence for the link between accreting LMXBs and
 isolated millisecond pulsars that had been anticipated previously \citep{alpar1982,radhakrishnan1984}.
 The discovery of a new class of transient, accretion-powered
 millisecond X-ray pulsars (AMSPs), simultaneously addressed both of
 these issues. They provide dramatic confirmation of the link between accreting
 LMXBs and millisecond pulsars \citep[e.g.][]{wijnands2006}: indeed, by combining these
 observations with measurements of burst oscillations and pulsations
 from other LMXBs, NS spin periods are now known for 20
 LMXBs. The reason why the pulsations are observed in this type of
 system, with a short orbital period, may be related to the low accretion
 rate (expected for such short periods), permitting accretion to the
 polar caps of the NS even in the presence of a relatively
 weak magnetic field.
 
 SAX~J1808.4$-$3658 is the prototypical AMSP. Initially detected by the
 \textit{BeppoSax} mission \citep{intzand1998}, subsequent \emph{Rossi X-ray Timing Explorer}
 (\emph{RXTE}) observations detected coherent millisecond pulsations at a frequency of 401~Hz from a
 NS in a $\sim$2 hour binary orbit. It was immediately
 realised that this was the long awaited missing link in the evolution of
 LMXBs to isolated millisecond pulsars \citep{wijnands1998,chakrabarty1998}. Subsequent
 observations revealed the optical counterpart at $R$ $\sim$16.1 mag \citep{roche1998}.
 \citet{homer2001} observed the optical counterpart in the quiescent state and found
 it to be at $R$ $\simeq 21$~mag.
 Using recent observations from the Gemini South telescope,
 \citet{deloye2008} measure average quiescent magnitudes of $g \simeq 21.7$~mag and $i \simeq 20.6$~mag.
 The 0.5 -- 10~keV X-ray luminosity ($L_{\mathrm{x}}$) of the system is $\sim$$2 \times 10^{36}$~erg~s$^{-1}$ in
 outburst \citep{intzand1998,intzand2001} and $5 \times 10^{31}$~erg~s$^{-1}$ in quiescence \citep{campana2002},
 for an assumed distance of 2.5 kpc \citep{intzand2001}.
 \citet{galloway2006} have calculated that the distance to the source is $\sim$3.5~kpc, assuming
 that the observed type-I X-ray bursts are helium-rich.
 
 The quiescent optical light curves exhibit sinusoidal variability, the phasing of which
 indicates that the optical maximum occurs when the secondary is directly
 behind the pulsar \citep{homer2001,deloye2008}. The observed
 optical flux is in excess of that expected assuming reprocessing of
 the measured X-ray flux.
 It was proposed that
 the irradiating flux was provided by the spin-down flux from the
 pulsar \citep{burderi2003,campana2004}, which was
 irradiating the face of the companion star, hence producing the observed
 modulation. This effect is similar to that observed in the binary
 millisecond pulsar PSR~B1957+20 \citep[e.g.][]{reynolds2007}.
 
 To account for their high spin frequencies, the neutron stars in AMSPs are expected to have
 accreted a few tenths of a solar mass during the lifetime of the binary, and hence should be
 systematically more massive than the canonical value of $\sim$1.4~M$_{\sun}$ \citep{thorsett1999}.
 Recent quiescent observations \citep{heinke2008} have hinted that the NS
 in SAX~J1808.4$-$3658 may be massive ($M_1 > 1.4$~M$_{\sun}$). Confirmation
 that the NS is indeed massive could have important implications
 for efforts to constrain the NS equation of state.
 
 For AMSPs, the orbital period ($P_{\mathrm{orb}}$) and projected semi-major axis of the primary
 orbit (and hence the projected primary velocity, $K_1$)
 are accurately known from analysis of the X-ray pulse arrival times. Therefore for these systems,
 only the projected secondary velocity ($K_2$) and the binary inclination angle ($i$) are required for
 the primary mass to be fully constrained.  Unfortunately,
 measurement of the radial velocity of the secondary star in SAX~J1808.4$-$3658 in quiescence
 is not feasible, because of the faintness of the secondary star.
 However, as \citet{steeghs2002} have shown in the case of the LMXB Scorpius X-1, it is possible to
 identify narrow Bowen blend emission line features \citep[e.g][]{mcclintock1975} formed
 in the secondary star by fluorescence of UV photons
 produced in the inner accretion disc, and use these to determine its radial velocity -- even when
 the binary is X-ray bright. This is likely the only means by which we will
 ever be able to measure the mass of the neutron stars in AMSPs, and confirm
 their expected massive nature.
 
 In addition, monitoring of these AMSPs provides us with an excellent
 opportunity to study accretion discs during outburst and to compare
 the observed structure with quantitative models. For example, in the
 short period, small mass ratio systems, the
 disc is expected to be elliptical and precessing \citep{whitehurst1988, haswell2001a}.
 Doppler tomography of these systems may provide evidence for such an
 elliptical disc. In addition, strongly X-ray irradiated discs are also
 susceptible to warping out of the orbital plane \citep*{pringle1996,foulkes2006}, which
 should also manifest itself spectroscopically. See \citet{elebert2008} for the first
 such study of the AMSP HETE~J1900.1$-$2455.
 
 In this paper, we present phased resolved optical
 spectroscopy and photometry of V4580 Sagittarii, the optical counterpart to the accreting X-ray
 millisecond pulsar SAX~J1808.4$-$3658,
 obtained during the 2008 outburst \citep{markwardt2008}. The paper is organised as
 follows: In \S2, we describe the spectroscopic and photometric
 observations. An analysis of these data is presented in \S3, in which we show how
 Doppler tomography of the Bowen blend \ion{N}{iii} $\lambda$4640.64 line allows us
 to measure the mass function of SAX~J1808.4$-$3658.
 Finally, we discuss these results in \S4, and present our conclusions in \S5.
 
 
 \section{Data}
 
 Our data consist of phase resolved spectroscopy and photometry of SAX
 J1808.4$-$3658, obtained during the 2008 September/October outburst.
 These observations were motivated by the need to obtain spectroscopy
 in the region of \ion{He}{ii} $\lambda$4686 and the Bowen blend near
 $\lambda$4640 in an effort to measure the radial velocity of the low
 mass secondary star \citep[$M_2 \sim$0.05 M$_{\sun}$;][]{bildsten2001},
 and hence place constraints on the mass of the
 pulsar. Although SAX~J1808.4$-$3658 is the brightest of all AMSPs in
 outburst, no optical spectra of the system in this state have been published to date.
 
\begin{figure}
    \begin{center}
      \includegraphics[scale=0.34, angle=-90]{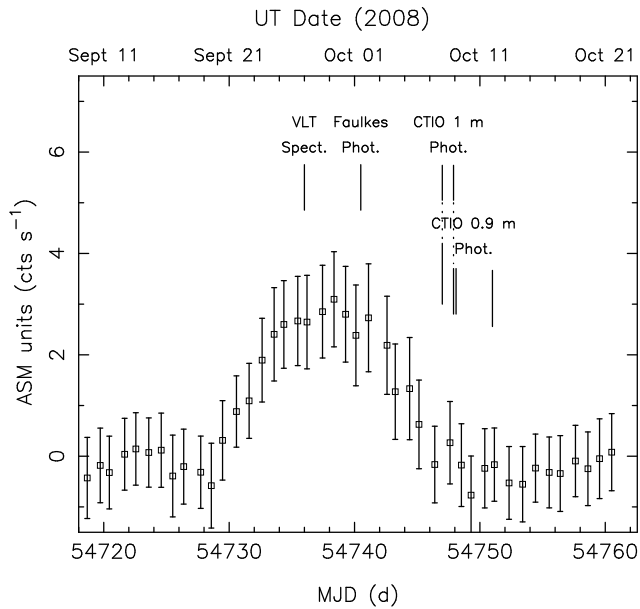}
      \caption{\emph{RXTE} ASM light curve, with the dates of our optical observations marked.}
      \label{asm}
    \end{center}
 \end{figure}

 Fig. \ref{asm} shows the \emph{RXTE} All-Sky Monitor (ASM) light curve
 of the outburst, with the dates of our observations marked. The
 observations at CTIO 1.0-m and 0.9-m telescopes on \mjd\ 54748 were
 taken at the same time, but have been offset slightly here for
 clarity.
 
 \subsection{Spectroscopy}
 \label{spect_data}
 
 A total of 39 optical spectra of SAX~J1808.4$-$3658 were acquired with
 the FOcal Reducer and low dispersion Spectrograph (FORS) 1 and FORS2
 spectrographs \citep{appenzeller1998}, mounted at the Cassegrain foci
 of the 8.2-m Kueyen (UT2) and Antu (UT1) Very Large Telescopes,
 respectively, at Cerro Paranal, Chile. The detector on FORS1 consists
 of a mosaic of two 2000 $\times$ 4000 pixel E2V CCDs (with 15~$\umu$m
 pixels) while FORS2 has a mosaic of two 2000 $\times$ 4000 pixel MIT
 CCDs (also with 15~$\umu$m pixels). The output data were binned by a
 factor of 2 in both spatial and dispersion directions.
 
 On 2008 September 27 \ut, 16 spectra were obtained using the 1200B grism
 on FORS1 with a 0.7~arcsec slit (seeing 0.8 -- 1.1~arcsec)
 and 360~s exposure times, while a
 further 23 spectra were obtained using the 600B grism on FORS2 with a
 1 arcsec slit (seeing 0.6 -- 1.0~arcsec) and 240~s exposure times. In all observations, the
 SAX~J1808.4$-$3658 spectrum was on CHIP1 of the FORS1 and FORS2
 mosaics. Appropriate bias, flat and comparison HgCd arc lamp frames
 were also taken.
 
 The data reduction process was identical for both the FORS1 and FORS2
 data. The frames were firstly processed using the
 \iraf\footnote{\iraf\ is distributed by the National Optical Astronomy
 Observatories, which are operated by the Association of Universities
 for Research in Astronomy, Inc., under cooperative agreement with the
 National Science Foundation.} \ccdproc\ routines to remove
 instrumental effects.  The spectra were then optimally extracted using
 tasks in the \iraf\ \kpnoslit\ package. Wavelength solutions for FORS1
 (FORS2) were found by fitting 2nd order cubic splines to 16 (23) lines
 in the arc spectra, giving RMS errors of $\sim$0.016 (0.036) \AA, and
 these solutions were then applied to the science frames.  The
 resulting wavelength range was $\lambda\lambda$3620 -- 5050 (3330 -- 6350),
 with a dispersion of $\sim$0.70 (1.49)~\AA~pixel$^{-1}$ and a
 resolution (measured from arc lamp lines) of $\sim$1.6 -- 1.9 (4.3 -- 4.8)~\AA.
 
 Because only one arc lamp exposure was taken during the daytime for each dataset, night skylines were examined
 in the SAX~J1808.4$-$3658
 frames to correct for instrumental flexure.
 For both the FORS1 and FORS2 datasets, a similar correction
 routine was employed. Firstly, the relative offsets between the wavelength scales for each of the individual spectra
 were found by cross-correlating each of the sky spectra against the last spectrum in the set (preliminary
 template), using the \rvsao\ package in \iraf\ \citep{kurtz1998}. These offsets were used to shift each of
 the individual spectra onto the wavelength scale of the preliminary template, and the resulting spectra were
 averaged to produce a master sky
 template. The positions of the strongest skylines were examined in this master template, in order to compute
 an average absolute shift. The master template was shifted by this amount, so that all lines appeared at their
 correct wavelengths, and the final shifts required for each spectrum were found by cross-correlating the
 spectra against the shifted master template. These shifts were then applied to both the data and error bands
 of each spectrum. The required shifts for the FORS1 data were within $\pm$0.07~\AA\ for each spectrum. For the
 FORS2 data, the required shifts were larger, $\sim$0.5 -- 0.7~\AA.
 The reduction procedure was repeated and confirmed using the FORS pipeline
 software\footnote{http://www.eso.org/sci/data-processing/software/pipelines}
 
\begin{figure*}
    \begin{center}
      \includegraphics[scale=0.44, angle=-90]{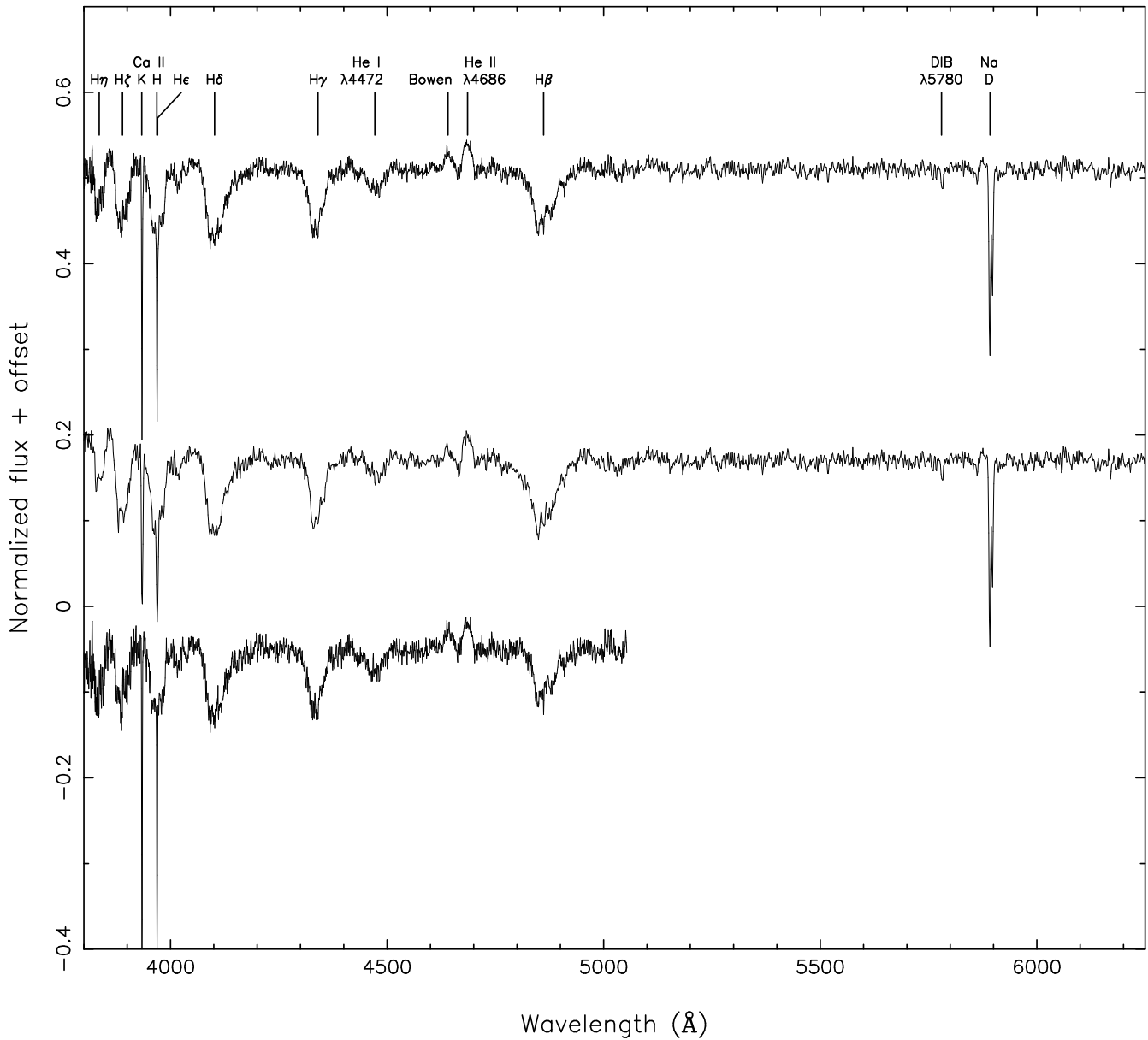}
      \caption{FORS1 (bottom), FORS2 (middle) and combined (top) averaged normalized spectra obtained on 2008 September 27.
               The spectra have been shifted to the rest frame of the system, using the systemic velocity of
               $-83$~km~s$^{-1}$ found in \S3, and offset vertically for clarity.}
      \label{norm1}
    \end{center}
 \end{figure*}

 The data were then imported into the \molly\ spectral analysis package, where they were shifted to the
 heliocentric rest frame, and re-binned onto a common velocity scale of 49~km~s$^{-1}$~pixel$^{-1}$ for the FORS1
 data and 95~km~s$^{-1}$~pixel$^{-1}$ for the FORS2 data. The continuum was removed by dividing by a cubic
 spline function (after masking the emission and absorption features) and subtracting unity. The normalized
 averaged FORS1 and FORS2 spectra are shown in Fig. \ref{norm1}. The orbital phase was determined using the
 X-ray ephemeris of \citet*{jain2008}.
 We transformed the $T_0$ (epoch when the pulsar is at the ascending node) of the ephemeris from \tdb\ to \ut,
 using the \xtime\ tool\footnote{http://heasarc.gsfc.nasa.gov/cgi-bin/Tools/xTime/xTime.pl},
 assuming that for
 the purposes of this work, \tdb\ is equal to \ttime\ and also that
 the heliocentre and barycentre are identical. We also shifted $T_0$ forward by one quarter of an orbital period,
 so that phase zero represents superior conjunction of the pulsar i.e. when the pulsar
 is at 90$\degr$ orbital longitude from the ascending node. The ephemeris we used was therefore
 $T_0 = 2450915.398669\ \hjd\ \ut$ and $P_{\mathrm{orb}} = 0.0839022785\ \mathrm{d}$.
 
 Because the primary lines of interest were the Bowen blend and
 \ion{He}{ii} $\lambda$4686, the data near these wavelengths were
 normalized separately, to ensure optimum continuum removal.
 
 \subsection{Photometry}
 
 In an effort to constrain the binary inclination angle via fitting the light curve
 with a physical model of the light sources
 \citep[e.g.][]{deloye2008}, we obtained
 phase resolved photometry of SAX~J1808.4$-$3658 from several
 telescopes during the 2008 September/October outburst. On 2008 October
 1 \ut, we obtained 20 Gunn/SDSS $i$-band images using the 2-m Faulkes
 South Telescope, at Siding Spring Observatory, Australia. The EA02
 camera was used, which has 2048 $\times$ 2048 pixels binned 2 $\times$
 2 into effectively 1024 $\times$ 1024 pixels, each of which is
 0.278~arcsec~pixel$^{-1}$; the total field of view is 4.7 $\times$ 4.7
 arcmin.

 On 2008 October 7/8 \ut\ we obtained 40 Gunn $r$-band images from the 1-m
 SMARTS/Yale Telescope, operated by the SMARTS consortium, at Cerro Tololo
 Inter-American Observatory (CTIO), Chile. A further 31 Gunn $r$-band images
 were obtained from the same telescope on 2008 October 8/9 \ut. The CCD system used for
 the observations was built by NOAO and Fermi National Accelerator
 Laboratory as part of the development of the Dark Energy Camera project
 \citep[see www.darkenergysurvey.org and][for more details]{depoy2008}. The
 detector in the system is a thick, fully depleted $2048 \times 2048$ pixel CCD with 15~$\umu$m
 pixels, which corresponded to approximately 0.3~arcsec~pixel$^{-1}$ on
 the sky. The exposure time for all observations was 300s. The raw images
 were bias corrected, trimmed, and flat-fielded using the \ccdproc\ routines
 in \iraf.
 
 On 2008 October 9 \ut\ we obtained 9 Bessel $V$-band images using the 0.9-m SMARTS Telescope, also at CTIO. On
 October 12, we acquired 8 Bessel $R$-band images. The
 SITe CCD has 2048 $\times$ 2046 pixels, with 0.4~arcsec~pixel$^{-1}$. As the CCD was read out using 4 amplifiers, the
 frames were bias corrected and flat-fielded using tasks in the \iraf\ \quadproc\ package.
 
\begin{figure}
    \begin{center}
      \includegraphics[scale=0.24, angle=-90]{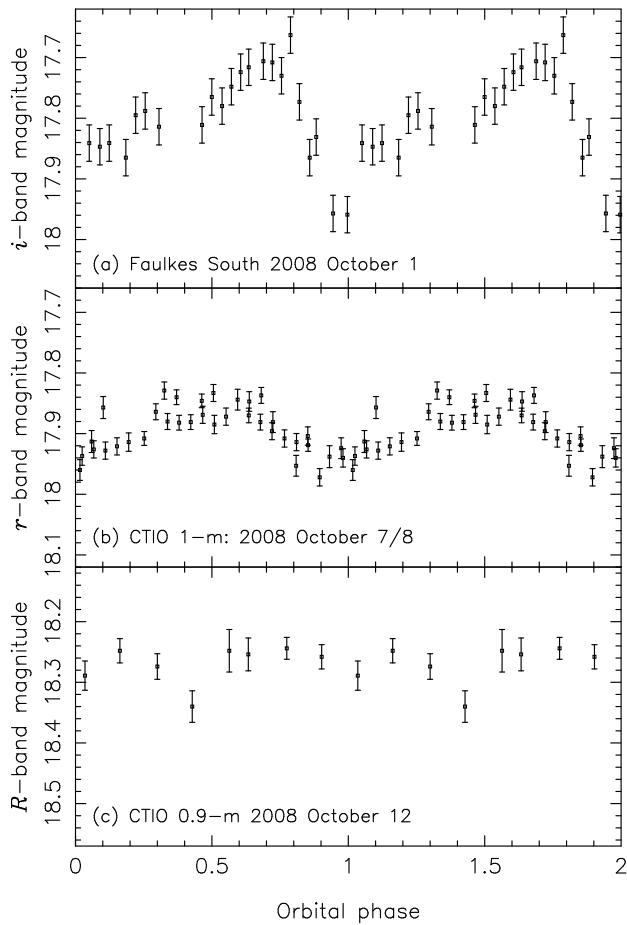}
      \caption{Light curves of SAX~J1808.4$-$3658, folded on the ephemeris defined in \S2.1, and plotted twice
               for clarity. (a) $i$-band data obtained with the Faulkes South telescope on 2008 October 1 \ut.
               (b) $r$-band data obtained with the CTIO 1-m telescope on 2008 October 7/8 \ut. (c) Bessel
               $R$-band data obtained with the 0.9-m CTIO telescope on 2008 October 12 \ut. All three light curves
               have the same magnitude scale.}
      \label{lc_phased}
    \end{center}
 \end{figure}

\begin{figure}
    \begin{center}
      \includegraphics[scale=0.24, angle=-90]{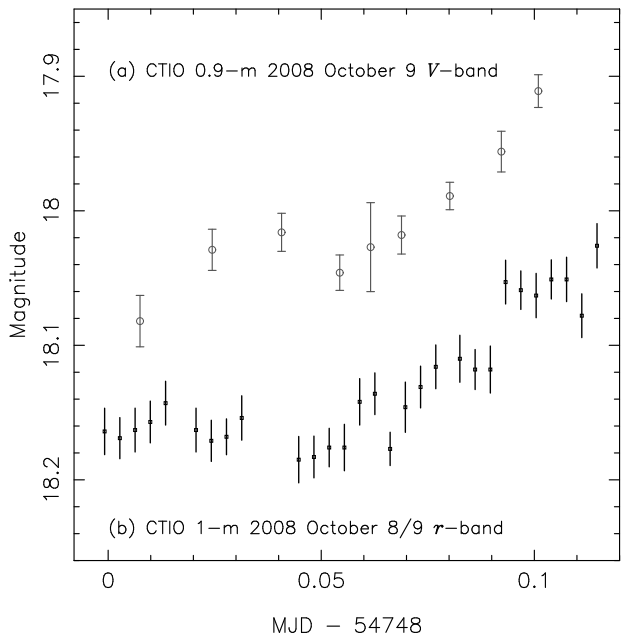}
      \caption{Light curves of SAX~J1808.4$-$3658.
               (a) $V$-band data obtained with the 0.9-m CTIO telescope on 2008 October 9 \ut\ (circles).
               (b) $r$-band data obtained with the CTIO 1-m telescope on 2008 October 8/9 \ut\ (squares). Note that the
               $r$-band light curve has been shifted down by 0.1~mag for clarity.}
      \label{lc_mjd}
    \end{center}
 \end{figure}

 Photometry was performed using the \daophot\ \citep{stetson1987} point spread function fitting package
 in \iraf.
 The magnitude of SAX~J1808.4$-$3658 was found by comparison with the three stars listed in table 1 of
 \citet*{greenhill2006}. For the $i$- and $r$-band frames, we firstly transformed the $R$-band magnitudes
 to $r$-band using equation 4 of \citet*{jordi2006} (for $V-R \leq 0.93$) and $I$-band to $i$-band using
 equation 8 of \citet{jordi2006}.
 Fig. \ref{lc_phased} shows three light curves, phased using the ephemeris defined in \S\ref{spect_data}, and
 plotted twice for clarity. Two of our data sets were obtained contemporaneously, and these are plotted in
 Fig. \ref{lc_mjd}.
 
 \section{Results}
 
 \subsection{Photometry}
 
 The main purpose of our photometry was to identify a modulation that
 could be used to constrain the binary inclination angle, by fitting
 the data using, for example, the \elc\ code of \citet{orosz2000}.
 
 Out first light curve was obtained by Faulkes South
 just as SAX~J1808.4$-$3658 was beginning to decline from the outburst.
 Although the minimum occurs
 near phase zero, the overall variation is more reminiscent of a
 superhump modulation \citep[see e.g.][]{patterson2005},
 or indeed a pre-eclipse hump from the hotspot
 \citep[see e.g.][]{grauer1994}, than a
 modulation due to the heated secondary star.
 Our second light curve was obtained when SAX~J1808.4$-$3658
 had declined further from
 outburst maximum: this light curve exhibits a modulation closest to what might
 expected from a heated secondary (see Fig. \ref{lc_phased}), although a superhump
 modulation cannot be ruled out.
 If this modulation
 is due to a heated secondary, a reasonable fit to the light curve (using \elc)
 can be achieved for any of the range of inclination angles found by \citet{deloye2008}.
 Our remaining light curves do not exhibit any obvious orbital
 modulation. Hence we conclude that these data cannot be used to place
 reliable constraints on the orbital inclination.
 Our light curves cover just over
 one orbital period each, so it also possible that the observed modulations are due to
 fluctuations in the mass accretion rate.
 
 {\citet{jordi2006} have very little data for $V - R \simeq 0$, but assuming
 that their transformations are valid in this regime,
 we transformed the 2008 October 7/8 \ut\ CTIO 1-m
 $r$-band magnitudes to Johnson-Cousins $R$-band by using
 the simultaneous $V$-band data, to give $V - R = 0.12 \pm 0.04$~mag.
 This is consistent with the measurements in fig. 2 of \citet{greenhill2006}.
 De-reddening the
 $R$- and $V$-band magnitudes using our measured $A_{\mathrm{V}} = 0.8 \pm 0.3$ (see \S\ref{spec_results}) yields
 $(V-R)_{0} = -0.08 \pm 0.09$~mag on this night.
 
 \subsection{Spectroscopy}
 \label{spec_results}
 
 Our spectra (Fig. \ref{norm1}) are dominated by broad Balmer
 absorption lines, with interstellar \ion{Ca}{ii} H and K and Na~D
 absorption lines, and emission lines from the Bowen blend near
 $\lambda$4640 and from \ion{He}{ii} $\lambda$4686. The diffuse
 interstellar band (DIB) at $\lambda$5780 is also present. The main
 characteristics of these features are summarised in Table
 \ref{ew_fwhm_table}.

 The Balmer lines from SAX~J1808.4$-$3658 itself extend to very high velocities, implying
 that they are produced in the optically thick accretion disc. There is
 some infilling of these lines by red-shifted emission features, most
 obviously in the H$\beta$ absorption line. The H$\gamma$ absorption line
 appears unusually narrow, compared with the other Balmer lines. It also appears to be blue-shifted
 relative to its rest wavelength, while the other Balmer lines seem to be relatively symmetrical
 about the rest wavelength, apart from the red-shifted emission. This may be due to higher levels
 of red-shifted emission in the H$\gamma$ line, although why this should be is unclear. This effect
 is present in both the FORS1 and FORS2 data, and so would not appear to be a wavelength calibration
 issue.

 \begin{table}
 \begin{center}
 \caption{Equivalent width and full-width half-maximum for the strongest features in the SAX~J1808.4$-$3658
           average spectrum.}
     \label{ew_fwhm_table}
     \begin{tabular}{| l | l | l | l |}
     \hline
                     & EW              &\multicolumn{2}{|c|}{\_\_\_\_\_\_ FWHM \_\_\_\_\_\_\ \ \ \ \ \ }\\
                     &(\AA)            &(\AA)& (km s$^{-1}$)\\
     \hline
     \ion{He}{ii} $\lambda$4686    & $-0.85 \pm 0.16$   & $23 \pm 6$ & $1500 \pm 400$   \\
     Bowen blend     & $-0.35 \pm 0.16$   & .......   & ........   \\
     H$\beta$        & $5.0 \pm 0.4$   & $70 \pm 6$  & $4300 \pm 400 $  \\
     H$\gamma$       & $2.8 \pm 0.3$   & $37 \pm 3$  & $2600 \pm 200 $  \\
     H$\delta$       & $4.5 \pm 0.3$   & $48 \pm 3$  & $3500 \pm 200 $  \\
     H$\epsilon$     & $2.5 \pm 0.2$   & $32 \pm 3$  & $2400 \pm 200 $  \\
     H$\zeta$        & $2.3 \pm 0.2$   & $29 \pm 3$  & $2200 \pm 200 $  \\
     H$\eta$         & $1.2 \pm 0.2$   & $22 \pm 4$  & $1700 \pm 300 $  \\
     \ion{He}{i} $\lambda$4472     & $1.0 \pm 0.3$   & $40 \pm 15$ & $2700 \pm 1000$   \\
     \hline
     \multicolumn{4}{|c|}{Interstellar features}\\
     \hline
     \ion{Ca}{ii} K $\lambda$3933.67  & $0.77 \pm 0.07$ & ........    & ........         \\
     \ion{Ca}{ii} H $\lambda$3968.47  & $0.56 \pm 0.08$ & ........    & ........         \\
     Na~D$_{1}$ $\lambda$5889.95 & $0.97 \pm 0.12$   & ........ & ........  \\
     Na~D$_{2}$ $\lambda$5895.92 & $0.68 \pm 0.12$   & ........ & ........   \\
     DIB $\lambda$5870 & $0.14 \pm 0.04$   & ........ & ........   \\
     \hline
     \end{tabular}
 \end{center}
 \end{table}
 
 \subsubsection{Extinction and distance}
 
 $E(B - V)$ is a function
 of the equivalent width (EW) of the DIB at $\lambda$5870. Using four
 of the stars in table 1 of \citet{jenniskens1994},
 $EW(5780)/E(B - V) = 0.52 \pm
 0.11$. For SAX~J1808.4$-$3658 $EW(5780) = 0.14 \pm 0.04$, therefore we find
 $E(B - V) = 0.27 \pm 0.11$. This correlation is also shown by
 \citet{webster1993} (fig. 5), and the point (0.27, 0.14) lies close
 to a best-fitting line through the low $E(B - V)$ ($<0.5$) points.
 The errors (here and in the remainder of the paper) represent the
 1$\sigma$ uncertainties (or 68.3\% confidence intervals for non-normal distributions),
 calculated using Monte Carlo simulations assuming Gaussian statistics.
 
 Fitting the X-ray spectra reveals that the hydrogen column density to
 the source ($N_{\mathrm{H}}$) is equal to $(1.3 \pm 0.1) \times 10^{21}$~cm$^{-2}$ \citep*{campana2008}. Taking
 $N_{\mathrm{H}}/E(B - V) = 5.0 \times 10^{21}$~cm$^{-2}$~mag$^{-1}$ \citep[][and
 references therein]{savage1979}, with a 1$\sigma$ error of 30\%
 \citep[][]{bohlin1978}, this implies $E(B - V) = 0.26^{+0.11}_{-0.06}$, consistent with
 the value found using the DIB.
 
 \citet{munari1997} examine the relationship between the EW of the Na~D$_1$
 line and $E(B - V)$. The values of $E(B - V)$ we calculate above are
 only consistent with their analysis if the Na~D absorption lines
 towards SAX~J1808.4$-$3658 have multiple components. Our spectral
 resolution is too low to determine if this is the case. The ratio of
 Na~D$_{1}$ to Na~D$_{2}$ is $\sim$1.4, which is also consistent with a
 relatively low $E(B - V)$.

 Taking $E(B - V) = 0.27 \pm 0.11$, $A_{\mathrm{V}} = 0.8 \pm 0.3$. This is
 consistent with the value found by \citet{wang2001}.
 
 \citet{megier2005} established relationships between the EW of the
 \ion{Ca}{ii} K and H interstellar absorption lines and the distance to
 the source. Such a relationship suggests that \ion{Ca}{ii} is homogeneously distributed
 throughout the interstellar medium \citep{galazutdinov2005}.
 These relationships suggest a distance to
 SAXJ1808.4$-$3658 of $2250 \pm 230$~pc based on the \ion{Ca}{ii} K
 line and $2700 \pm 400$~pc based on the \ion{Ca}{ii} H line.
 The combined distance estimate is $2500 \pm 400$~pc.
 
 \subsubsection{Projected velocity of the donor star}
 
 As discussed in \S\ref{intro}, for systems with low luminosity companions like
 SAX~J1808.4$-$3658, the only way to determine the projected velocity of
 the secondary star is by observing the radial velocity shifts of the
 Bowen emission lines near $\lambda$4640. The primary features
 contained in the Bowen blend are \ion{N}{iii} emission lines at
 $\lambda\lambda$4634.13, 4640.64, 4641.85 and 4641.96, and
 \ion{C}{iii} emission lines at $\lambda\lambda$4647.4 and 4650.1. In
 the majority of systems so far observed, the \ion{N}{iii}
 $\lambda$4640.64 line is the strongest of these \citep{cornelisse2008}.
 
 In some systems, the velocity shift of these individual lines can be
 observed as S-waves in a trailed spectrogram
 \citep[e.g][]{steeghs2002,casares2006}. In other cases, the S/N and
 resolution of the observations are such that the S-waves are not
 obvious, but the presence of velocity shifts is revealed by
 performing Doppler tomography \citep{marsh1988,marsh2001} of the spectra \citep[e.g.][]{casares2003}.
 \citet{casares2004} report on a successful implementation of this
 technique for the AMSP XTE~J1814$-$338, which revealed the velocity of
 the donor star emission to be 345~km~s$^{-1}$.  \citet{elebert2008}
 attempted this for another AMSP, HETE~J1900.1$-$2455, but the S/N of
 their data was insufficient to provide a reliable measurement.
 
 Before attempting to detect the secondary star in either the trailed
 spectra or Doppler tomograms, we firstly used the existing information
 about the system to determine the expected range of projected
 secondary velocities. Analysis of the pulse arrival times
 \citep{chakrabarty1998,jain2008} provides us with very accurate values
 of $P_{\mathrm{orb}}$ ($7249.156862 \pm 0.000005$~s) and
 $a_{1}\sin{i}$ ($62.809 \pm 0.001$~light-ms), which combine to give
 the primary velocity, $K_1 = 2\pi a_{1}\sin{i}/P_{\mathrm{orb}} =
 16.32056 \pm 0.00026$~km~s$^{-1}$. Since the primary is a NS,
 we assume an extreme primary mass range of 1.4 -- 3.0~M$_{\sun}$,
 and that $i$ lies between 5$\degr$ and 90$\degr$.
 
 The mass function equation relates the mass of the primary to these
 observable quantities.
 
 \begin{equation}
      f(M) = \frac{M_{1}\sin^{3}{i}}{(1 + q)^{2}} = \frac{K_{2}^{3}P_{\mathrm{orb}}}{2\pi G} = \frac{K_{1}^{3}P_{\mathrm{orb}}}{2\pi Gq^{3}}
      \label{mass_function}
 \end{equation}
 
 \noindent
 where $q$ is the binary mass ratio ($\equiv M_{2}/M_{1} = K_{1}/K_{2}$)
 By solving Equation \ref{mass_function} for the
 cubic in $q$, we calculated $K_2$ for a range of $i$ and $M_1$, and
 plot $K_2$ vs. $M_1$ for several values of $i$ in Fig. \ref{kvm}.
 \begin{figure}
    \begin{center}
      \includegraphics[scale=0.49, angle=-90]{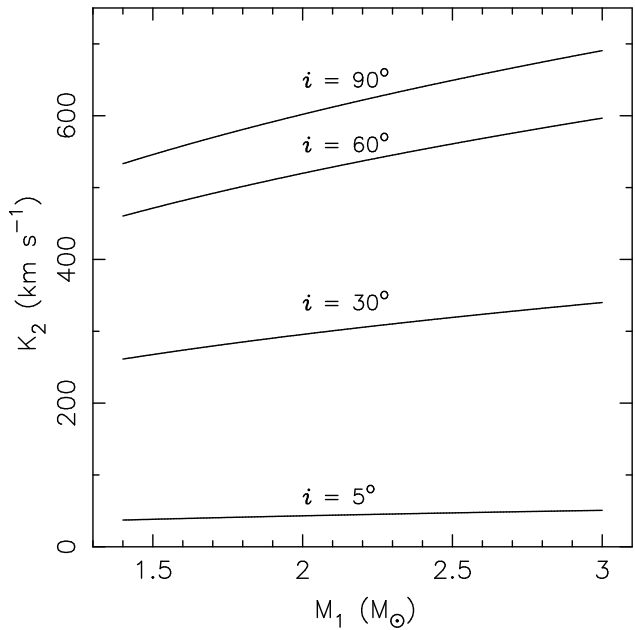}
      \caption{$K_2$ vs. $M_1$ for inclination angles of 5$\degr$, 30$\degr$, 60$\degr$ and 90$\degr$.}
      \label{kvm}
    \end{center}
 \end{figure}
 This plot shows that, for $M_1 < 3$~M$_{\sun}$ and all possible
 inclination angles, $K_2 < 700$~km~s$^{-1}$. Because the ephemeris is
 known so accurately from the pulse timing analysis, 
 emission from the secondary should appear very close to the
 positive $V_{\mathrm{y}}$ axis, and at a velocity less than 700~km~s$^{-1}$, in
 our Doppler tomograms.
 
 In order to perform Doppler tomography, a parameter which must be
 known is the systemic velocity ($\gamma$), as this determines the offset
 of any underlying S-wave from the rest wavelength.
 Ideally, this should be obtained using radial velocity studies of the
 secondary star: this is impossible here however. Instead, we 
 determined a value for $\gamma$ by fitting a Gaussian to the \ion{He}{ii}
 $\lambda$4686 line, after masking the line core. If the flux in the
 wings of this emission line originates in the inner accretion disc,
 then a fit to the wings of this line should provide a reasonable estimate of
 $\gamma$, where the flux is less contaminated by
 irregularities in the
 outer disc such as the gas stream/accretion disc impact region
 \citep[although see the discussion by][]{marsh1998}.
 For the NS LMXB Cygnus X-2 \citep{elebert2009}, and BH
 LMXB XTE~J1118+480 (Elebert et~al., in preparation.), the wings of this line,
 averaged over an entire orbital period, 
 provide values for $\gamma$ consistent (within the errors) with those found
 by analysis of the secondary star absorption line spectra.
 
 For the FORS1 data, the best fit gives $\gamma = -77 \pm 14$~km~s$^{-1}$,
 while for the FORS2 data, we get a best fit $\gamma$ of $-77 \pm
 16$~km~s$^{-1}$. Unfortunately, the Balmer absorption lines were too
 broad to confirm these values.
 
 Our Doppler tomograms were constructed using the maximum entropy
 method (MEM), as implemented in \doppler. An initial value for
 $\gamma$ of $-77$~km~s$^{-1}$ was chosen for both the FORS1 and FORS2
 tomograms. Since the \ion{N}{iii} $\lambda$4640.64 emission line is
 generally the strongest in the Bowen blend, we first attempted to make
 Doppler tomograms of this line. The tomograms were iterated to
 minimize the reduced $\chi^2$ ($\chi^{2}_{\nu}$) between the data and tomogram, while maintaining
 a smooth image.
 Both FORS1 and FORS2 tomograms showed clear emission near
 the positive $V_{\mathrm{y}}$ axis,
 where emission from the secondary is expected, and in both cases this
 emission was at a value of $\sim$300~km~s$^{-1}$.
 Using the 3$\sigma$ range in systemic velocity,
 Doppler tomograms of the other
 Bowen blend lines did not show any emission from near the $V_{\mathrm{y}}$ axis,
 nor any focused spots of emission at any phase or velocity.
 
 We then combined both datasets and a Gaussian fit to the wings of the
 averaged \ion{He}{ii} $\lambda$4686 line gave a value for $\gamma$ of
 $-83 \pm 12$~km~s$^{-1}$. 
 As an independent check on the value of $\gamma$, Doppler
 tomograms of the \ion{N}{iii} $\lambda$4640.64 emission line were created
 for a range of $\gamma$ values, and
 the spot characteristics (peak and FWHM) were measured in each tomogram by
 fitting with a point spread function. When the ratio of the peak to FWHM was
 plotted against $\gamma$, a maximum was observed at $\gamma = -55 \pm 20$~km~s$^{-1}$,
 which we regard as consistent with our estimate from the \ion{He}{ii}
 $\lambda$4686 line. We note that the spot of emission is closest to the
 $V_{\mathrm{y}}$ axis near $\gamma = -80$~km~s$^{-1}$, and so we adopt a value
 for $\gamma$ of $-83 \pm 12$~km~s$^{-1}$, as determined from the \ion{He}{ii}
 $\lambda$4686 line fitting.
 
 We created several Doppler tomograms from these data using
 $\gamma = -83$~km~s$^{-1}$, with different velocity binnings, phase binnings and
 filtering. In all cases, we observed a bright spot of emission on the
 positive $V_{\mathrm{y}}$ axis near 300~km~s$^{-1}$.
 Fig. \ref{dt_all} shows the
 Doppler tomogram for the case where we re-binnned all the data at
 49~km~s$^{-1}$, and into 20 phase bins. 
 The centroid of the spot of emission is at a velocity of
 $K_{\mathrm{em}} = 324 \pm 8$~km~s$^{-1}$. For Doppler tomograms within
 the 1$\sigma$ range of $\gamma$ values, the spot remained focused and close to the same
 value of $324 \pm 8$~km~s$^{-1}$. We estimate that the effect of the spot location
 dependency on $\gamma$ increases the 1$\sigma$ error on
 $K_{\mathrm{em}}$ to 15~km~s$^{-1}$. Again, none of the other Bowen
 blend lines showed any focused spots of emission in their tomograms.
 The spectra extracted using the FORS pipeline software, and processed in
 an identical manner, yielded identical results, within the uncertainties.
 
 \begin{figure}
    \begin{center}
      \includegraphics[scale=0.42, angle=-90]{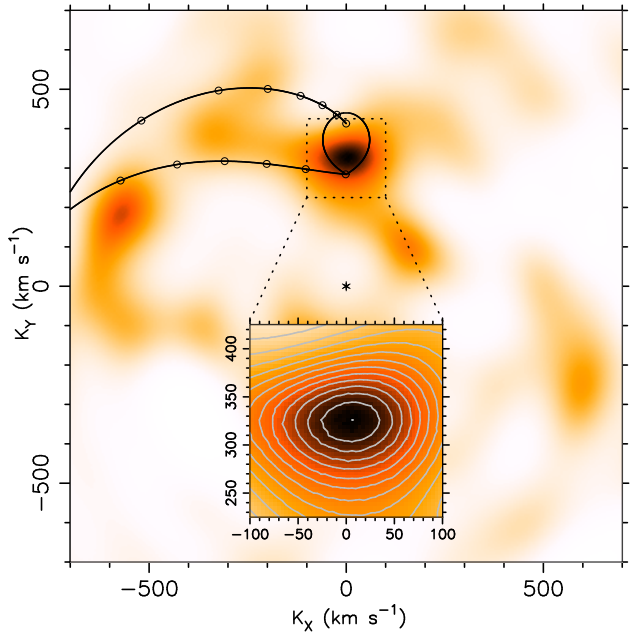}
      \caption{Doppler tomogram of the \ion{N}{iii} $\lambda$4640.64 emission line using the combined FORS1 and FORS2
               data. The Roche lobe of the secondary for $K_2 = 370$~km~s$^{-1}$ and $q = 0.044$, the gas stream from
               the inner Lagrangian point and the velocity of the disc along the gas stream are overplotted.}
      \label{dt_all}
    \end{center}
 \end{figure}
 
 \begin{figure}
    \begin{center}
      \includegraphics[scale=0.41, angle=-90]{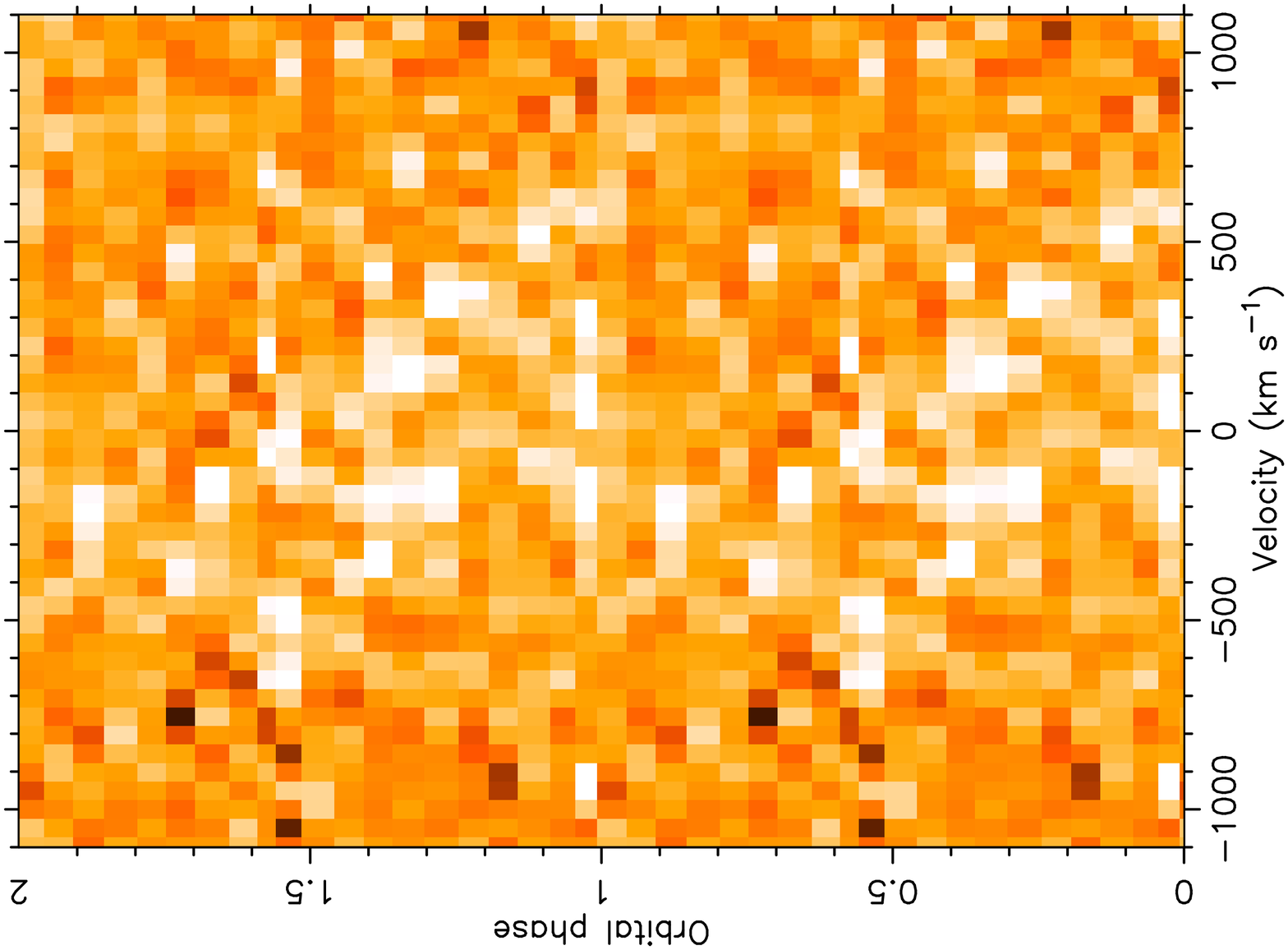}
      \includegraphics[scale=0.41, angle=-90]{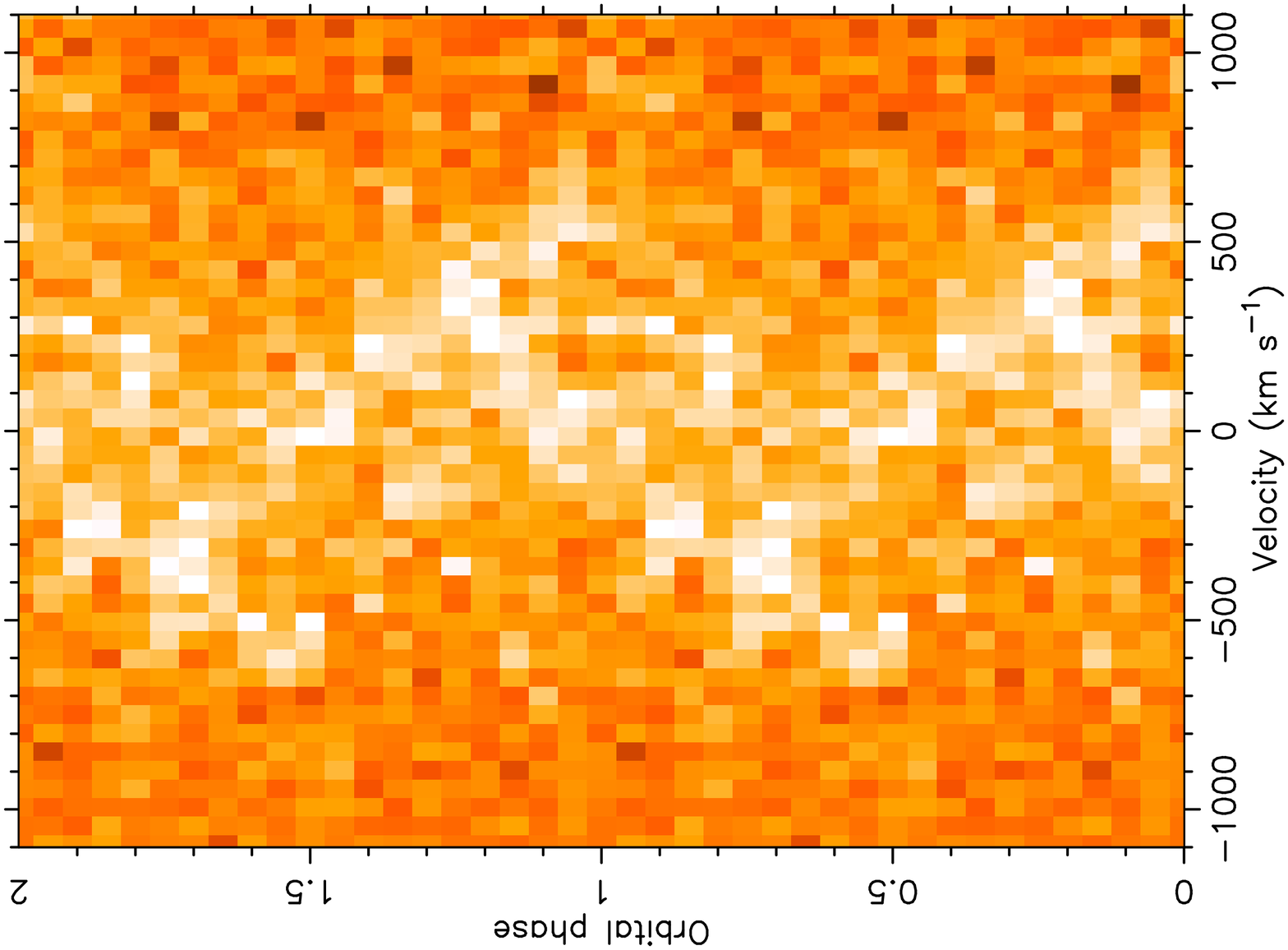}
      \caption{Trailed spectrum of the \ion{N}{iii} $\lambda$4640.64 emission line using the combined FORS1 and
               FORS2 data, in 20 phase bins (top panel) and
               trailed spectrum computed from the Doppler tomogram in Fig. \ref{dt_all} (bottom panel).}
      \label{trailed_all}
    \end{center}
 \end{figure}

 \subsubsection{The $K$-correction}
 
 $K_{\mathrm{em}}$ is the velocity of the inner hemisphere of the
 secondary, and a correction must be applied to obtain the velocity of
 the centre of mass of the secondary -- the $K$-correction. This
 correction is the ratio of the velocity of the Bowen emitting region
 to the velocity of the centre of mass of the secondary, and depends on
 how the secondary is irradiated by the central source. By simulating
 the irradiation of the secondary, \citet*{munoz-darias2005} computed
 $K$-correction values for a range of mass ratios, inclinations and
 disc flaring angles (the opening angle of the disc rim above the
 plane, $\alpha$).
 They fit their simulation results with 4th order polynomials in $q$, and present
 the coefficients for various values of $\alpha$, and for $i = 40\degr$ and $90\degr$.
 
 Since $K_1$ is known to high accuracy from the pulse timing analysis,
 we estimated a value of $q$ as the ratio $K_1$ to $K_{\mathrm{em}}$,
 calculated a $K$-correction based on this, then iterated until a
 value of $K_2$ was converged upon.  We used the polynomial
 coefficients computed for $i=40\degr$ (although we found very little
 difference for the $i = 90\degr$ coefficients), and calculated the
 $K$-correction for a range of values of $\alpha$. The maximum value of
 $\alpha$, for which irradiation of the secondary would still occur,
 was also calculated for the different values of $q$.  Based on this,
 $\alpha$ must be less than $\sim$9$\degr$, since, regardless of the
 value of $q$, any value of $\alpha$ greater than 9$\degr$ would cause
 the secondary to be totally shielded from the primary/inner disc by
 the disc. Within the 1$\sigma$ range of values for $K_{\mathrm{em}}$
 (309 -- 339~km~s$^{-1}$) and for all values of $\alpha \leq 9\degr$,
 we find $K_2 = 370 \pm 40$~km~s$^{-1}$.
 Therefore, our estimate for $K_2$ is quite conservative, as it takes the full range
 of possible values of $\alpha$
 into account, since we have no way of determining $\alpha$.
 Combining our measurement of $K_2 = 370 \pm 40$~km~s$^{-1}$ with the existing
 parameters and Equation \ref{mass_function}, we find $q = 0.044^{+0.005}_{-0.004}$,
 $f(M) = 0.44^{+0.16}_{-0.13} $~M$_{\sun}$
 and $M_{1}\sin^{3}{i} = 0.48^{+0.17}_{-0.14}$~M$_{\sun}$.
 
 \subsubsection{He {\small II} $\lambda$4686 Doppler tomography}
 
 Using $\gamma = -83$~km~s$^{-1}$ we created a Doppler tomogram of the
 \ion{He}{ii} $\lambda$4686 emission line.  The input data we used was
 similar to that used for the \ion{N}{iii} $\lambda$4640.64 line -- velocity binned to
 49~km~s$^{-1}$~pixel$^{-1}$, in 20 phase bins.
 Fig. \ref{dt_4686} shows the Doppler tomogram of the
 \ion{He}{ii} $\lambda$4686 line, with the Roche lobe of the secondary,
 gas stream velocity and velocity of the stream along the accretion
 plotted for $K_2 = 370$~km~s$^{-1}$ and $q = 0.044$.
 The main feature of the tomogram is a bright spot of emission, along the gas stream
 trajectory. Whether this coincides with the location of the gas stream/accretion disc
 impact point is unclear. There is also a roughly circular pattern, due to the
 accretion disc, although this is much more apparent in tomograms where the $\chi^{2}_{\nu}$ has
 not been decreased to as low a value.
 Fig. \ref{trailed_4686} shows the trailed spectrogram used
 to produce this Doppler tomogram, as well as the spectrogram computed from the Doppler
 tomogram.
 
 \begin{figure}
    \begin{center}
      \includegraphics[scale=0.44, angle=-90]{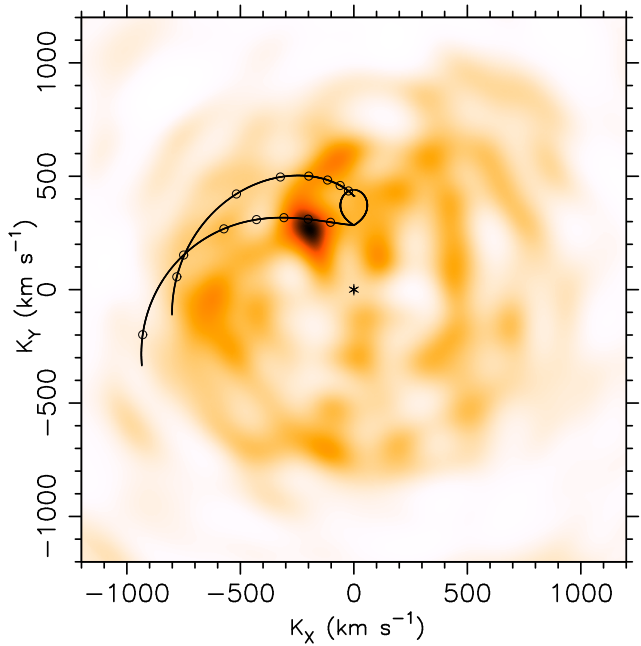}
      \caption{Doppler tomogram of the combined FORS1 and FORS2 data for the \ion{He}{ii} $\lambda$4686 emission line.
               The Roche lobe of
               the secondary for $K_2 = 370$~km~s$^{-1}$ and $q = 0.044$, gas stream from the inner Lagrangian point and
               velocity of the disc along the gas stream are overplotted.}
      \label{dt_4686}
    \end{center}
 \end{figure}
 
 \begin{figure}
    \begin{center}
      \includegraphics[scale=0.41, angle=-90]{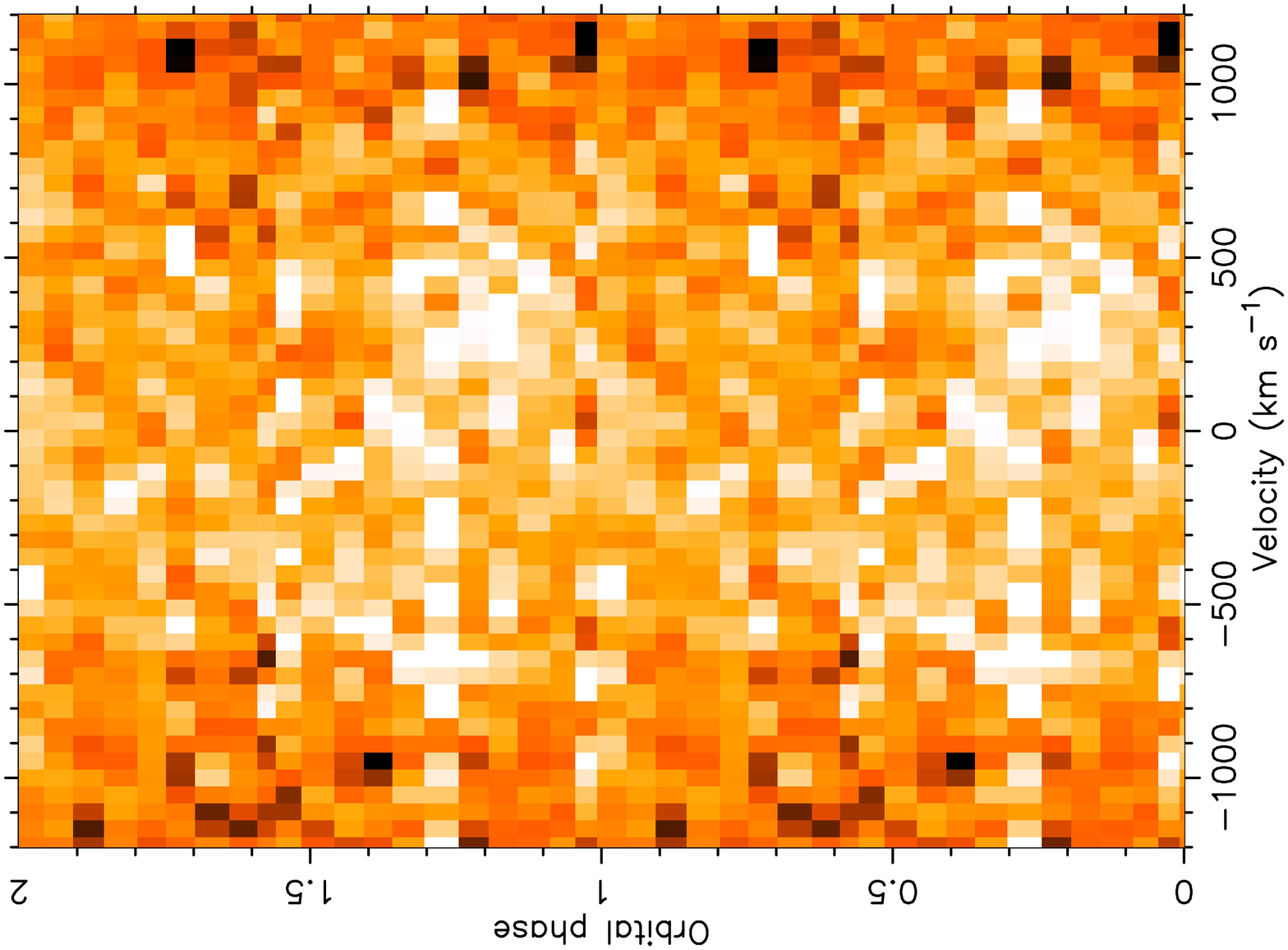}
      \includegraphics[scale=0.41, angle=-90]{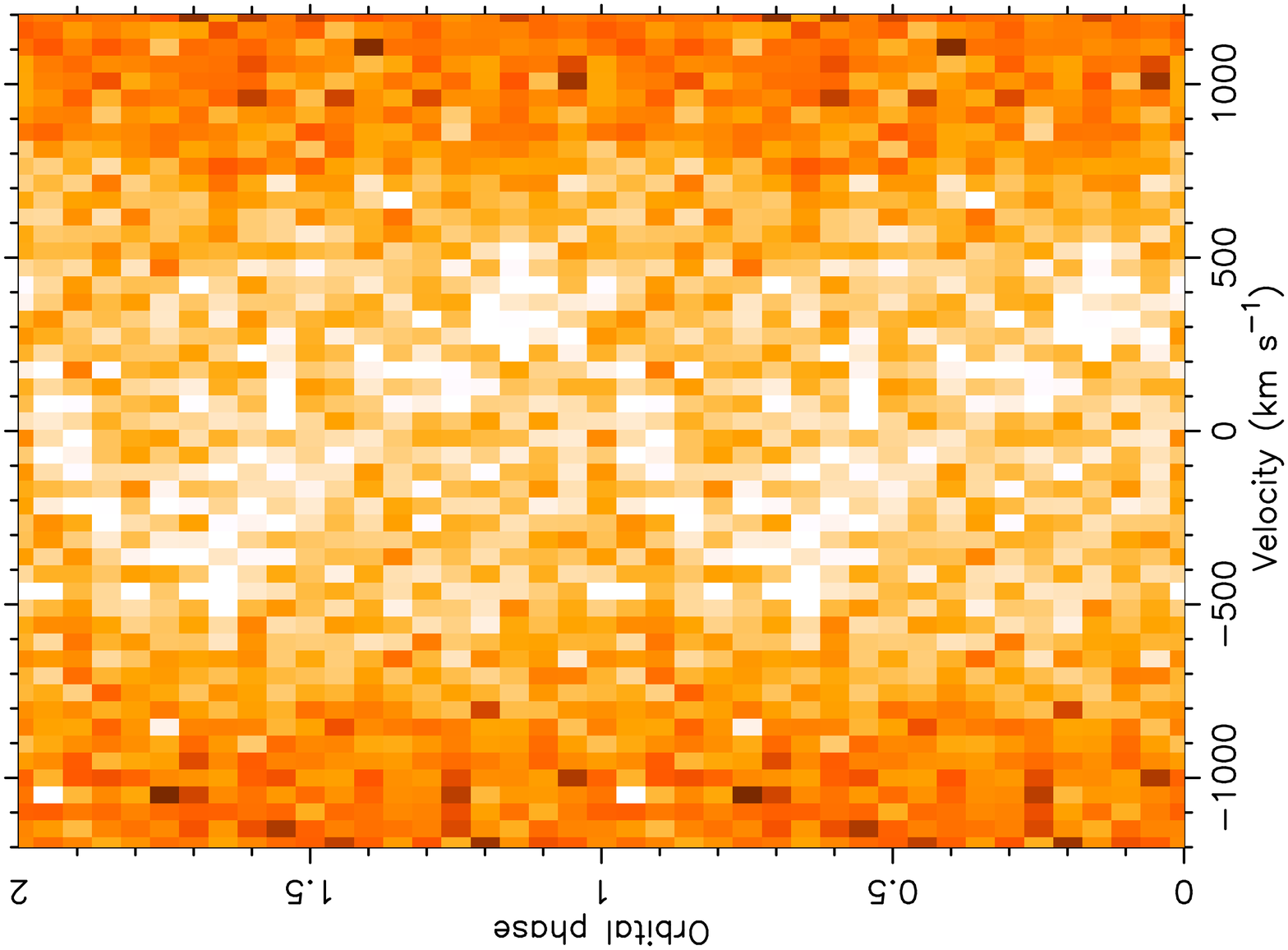}
      \caption{Trailed spectrum of the \ion{He}{ii} $\lambda$4686 emission line using the combined FORS1 and
               FORS2 data, in 20 phase bins (top panel) and
               trailed spectrum computed from the Doppler tomogram in Fig. \ref{dt_4686} (bottom panel).}
      \label{trailed_4686}
    \end{center}
 \end{figure}

 
 \section{Discussion}
 
 \subsection{The mass of the pulsar}
 
 Fig. \ref{kvi} shows a plot of $K_2$ vs. $i$ for
 primary masses of 1.4, 1.8, 2.2, 2.6 and 3.0~M$_{\sun}$. Based on our
 estimates for the system parameters, we calculate that for a primary
 mass between 1.4 and 3.0~M$_{\sun}$, the inclination angle must lie
 between 33$\degr$ and 44$\degr$ for $K_2 = 370$~km~s$^{-1}$ and between
 29$\degr$ and 51$\degr$ for the 1$\sigma$ range of values in $K_2$.
 For pulsar masses of 1.4 -- 3.0~M$_{\sun}$, our value of $q$ leads to companion
 masses of $\sim$0.06 -- 0.13~M$_{\sun}$.
 
 \begin{figure}
    \begin{center}
      \includegraphics[scale=0.48, angle=-90]{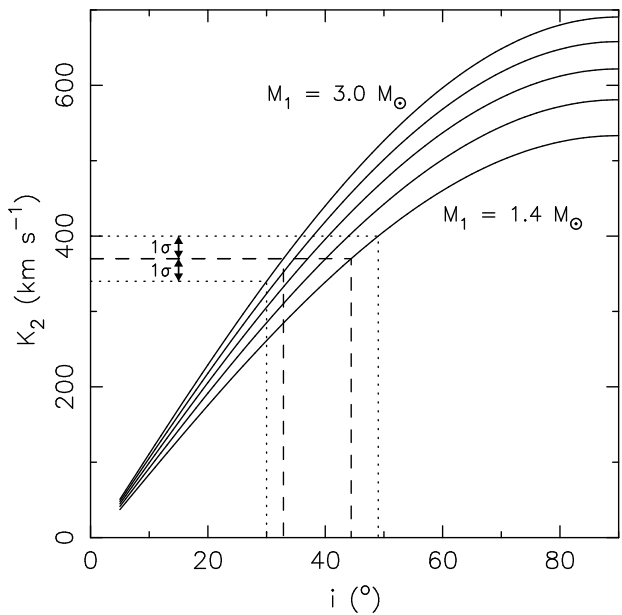}
      \caption{$K_2$ vs. $i$ for primary masses of 1.4, 1.8, 2.2, 2.6 and
       3.0~M$_{\sun}$. The dashed lines mark the allowed range of inclination angles
       if $K_2 = 370$~km~s$^{-1}$, and the dotted lines show the range of angles at
       the 1$\sigma$ limits on $K_2$.}
      \label{kvi}
    \end{center}
 \end{figure}

 Our data provide us with a value for $M_{1}\sin^{3}{i} = 0.48^{+0.17}_{-0.14}$~M$_{\sun}$.
 In order to determine the primary mass, the only remaining parameter that must be found is the
 inclination angle.
 There have been several attempts to constrain the orbital inclination of
 SAX~J1808.4$-$3658.
 \citet{deloye2008} have published
 Gemini SDSS $i$- and $g$-band light curves of SAX~J1808.4$-$3658 in
 quiescence, and find that the 2$\sigma$ range of values for
 the inclination is 32 -- 74$\degr$, irrespective of the primary mass. Hence,
 for an inclination of $53\degr \pm 11\degr$ (1$\sigma$),
 we derive a pulsar mass of $1.0^{+0.8}_{-0.4}$~M$_{\sun}$.

 X-ray measurements taken during the 2008 outburst also help to
 constrain the inclination angle. For example, using observations from \emph{Suzaku},
 \citet{cackett2009} fit the relativistically
 broadened Fe K line, and found a value of $i = 55^{+8}_{-4}$~degrees (90\% confidence).
 Combining this estimate with our measurements leads to a primary mass of
 $0.9 \pm 0.3$~M$_{\sun}$ (1$\sigma$).
 By analysing \emph{RXTE} data from the 2002 outburst, \citet{ibragimov2009} find a value
 of $i = 60 \pm 5\degr$, consistent with the results of
 \citet{cackett2009} and \citet{deloye2008}, within the uncertainties.

 Although all of these estimates are somewhat model dependent, it is clear
 that, when combined with our results, they do not support a heavy NS in SAX~J1808.4$-$3658.
 This would suggest that the NS in SAX~J1808.4$-$3658 has not accreted a significant amount
 of material, implying that the mass transfer has been non-conservative over the lifetime
 of the system.
 
 \subsection{Accretion disc}
 
 Our spectra show broad Balmer absorption lines, with some in-filling
 of the cores on the red sides of the lines. These absorption lines originate in
 the optically thick accretion disc. The spectra show similarities to
 the spectra of KS Ursae Majoris \citep{zhao2006}, an SU UMa type
 (i.e. exhibits superhumps during superoutbursts) cataclysmic variable
 system. In particular, the Balmer absorption lines show some infilling
 in both cases, with this effect more pronounced at longer wavelengths.
 The offset emission we observe may have an origin in an eccentric precessing accretion disc,
 as was also observed in the spectra of the LMXB XTE~J1118+480 \citep{torres2002}.
 Helium absorption is at most very
 weak, which suggests that the accretion disc is dominated by
 hydrogen.
 
 The optical photometry reveals a modulation which may be due to the
 phase dependent visibility of the irradiated face of the secondary
 star, similar to the situation in quiescence
 \citep[e.g.][]{deloye2008}.  However, it is more likely that this
 modulation is due to a superhumping accretion disc, or simply due
 to variations in the mass transfer rate onto the NS. Superhumps are
 periodic optical modulations, originally seen during superoutbursts of
 SU UMa dwarf novae (DNe), but also seen in some LMXBs in outburst
 \citep[][and references therein]{haswell2001a}.  The superhump period
 ($P_{\mathrm{sh}}$) is typically a few percent longer than
 $P_{\mathrm{orb}}$, although we have insufficient data here to
 determine if this is the case for SAX~J1808.4$-$3658.

 The Faulkes $i$-band light curve shown in Fig. \ref{lc_phased}(a) is
 more like a sawtooth pattern, rather than the sinusoid-like variation
 expected if the modulation was due to the secondary star. Although the
 minimum of this light curve occurs near phase zero, the maximum
 brightness is at phase 0.7 -- 0.8, rather than phase 0.5.
 The CTIO 1-m $r$-band light curve in Fig. \ref{lc_phased}(b) is more
 sinusoidal, but the amplitude of the modulation is a factor of $\sim$2
 lower than in the $i$-band light curve. The CTIO 0.9-m $R$-band light
 curve shows no discernible modulation.
 Another AMSP, HETE~J1900.1$-$2455
 exhibits superhumps \citep{elebert2008}, and in this case the
 amplitude of the superhumps varied by a factor of $\sim$3 over 11
 days, without any significant change in the X-ray brightness. For
 $L_{\mathrm{x}}$ $\sim$10$^{36}$~erg~s$^{-1}$ the accretion disc is also likely to
 exhibit warping caused by the irradiation.
 Whether or not this is the case for SAX~J1808.4$-$3658 is unclear, but
 the combination of apsidal and warped precession may explain why some
 of our light curves appear to exhibit periodic variations, and others do not.
 We also note that most of our light curves 
 cover a single orbital period only, and hence we cannot exclude the 
 possibility that purely random variations in the mass transfer rate 
 could give rise to the apparent superhump variability discussed here.
 
 
 \section{Conclusions}
 
 We have presented Doppler tomography of the \ion{N}{iii}
 $\lambda$4640.64 emission line for SAX~J1808.4$-$3658, from which we find the
 velocity of the emission to be at $324 \pm 15$~km~s$^{-1}$. Applying the
 $K$-correction, we find the projected velocity of the centre of mass of the
 secondary star to be $370 \pm 40$~km~s$^{-1}$. Combining this with
 existing parameters, we constrain the mass ratio to be
 $0.044^{+0.005}_{-0.004}$, and the mass function of the pulsar to be
 $0.44^{+0.16}_{-0.13}$~M$_{\sun}$. Hence, the mass of the pulsar is
 $0.48^{+0.17}_{-0.14} / \sin^{3}{i}$~M$_{\sun}$.
 
 Using various estimates of the binary inclination angle we find no evidence
 to suggest that the neutron star in SAX~J1808.4$-$3658 is more massive than the
 canonical value of 1.4~M$_{\sun}$, despite the fact that AMSPs are expected to
 have accreted a significant fraction of a solar mass in order to attain
 such rapid spin frequencies.

 An important next step in determining the mass of the neutron star in SAX~J1808.4$-$3658 is to
 confirm the orbital inclination estimate from the Fe K line fitting \citep{cackett2009}. The outburst
 optical light curves appear to be dominated by the disc and not, for example,
 by an X-ray heated secondary that could be modelled to infer the 
 inclination. In quiescence, however, a modulation due to a heated
 secondary is clearly visible \citep[e.g.][]{deloye2008}, and more detailed
 multicolour modelling here is probably the best route to confirming the
 inclination.

 Determining the masses of the neutron star in AMSPs is challenging for
 a number of reasons. However, because of the ability to accurately
 measure the orbital period and projected velocity of the neutron star using
 the X-ray pulses, these AMSPs remain the most promising systems for
 determining neutron star masses, helping to constrain the
 equation of state which describes matter in these neturon stars.

 
 \section*{Acknowledgments}
 
 Part of this work is based on observations made at the European Southern Observatory,
 Chile.
 We thank the ESO Director General for a generous allocation of Director's
 Discretionary Time (DDT 281.D-5060, 281.D-5061).
 The Faulkes Telescope Project is an educational and research arm of the Las Cumbres Observatory Global Telescope
 Network (LCOGTN). We thank the staff and students of Glenlola Collegiate, South Downs Planetarium, Oundle School,
 Dartford Grammar School and Portsmouth Grammar School for performing some of the Faulkes Telescope observations.
 Thanks to C.~Izzo and S.~Bagnulo for advice on applying the skyline correction to our spectra.
 This research made use of NASA's Astrophysics Data System, and the SIMBAD database, operated at CDS, Strasbourg, France.
 We thank J. A. Orosz for use of the \elc\ code. We acknowledge the use of \molly\ and \doppler\ software
 packages developed by T.~R.~Marsh, University of Warwick.
 X-ray quick-look results provided by the ASM/RXTE team.

 We thank Ricardo Schmidt and Marco Bonati of CTIO for building the Dark Energy Camera CCD
 system and Juan Estrada and the entire CCD production effort at Fermilab
 for creating the CCD detector. Fermilab is operated by the Fermi Research
 Alliance, LLC under contract no. DE-AC02-07CH11359 with the United States
 Department of Energy.

 PE and PJC acknowledge support from Science Foundation Ireland.
 FL would like to acknowledge support from the Dill Faulkes Educational Trust.

\appendix
 
\bsp
 
\label{lastpage}
 
\end{document}